\documentclass[twoside,twocolumn,9pt]{article}
\usepackage{extsizes}
\usepackage[super,sort&compress,comma]{natbib} 
\usepackage[left=1.5cm, right=1.5cm, top=1.785cm, bottom=2.0cm]{geometry}
\usepackage{balance}
\usepackage{mathptmx}
\usepackage{sectsty}
\usepackage{graphicx} 
\usepackage{lastpage}
\usepackage[format=plain,justification=justified,singlelinecheck=false,font={stretch=1.125,small,sf},labelfont=bf,labelsep=space]{caption}
\usepackage{float}
\usepackage{fancyhdr}
\usepackage{fnpos}
\usepackage[version=4]{mhchem}

\usepackage{chemfig}
\usepackage[english]{babel}
\addto{\captionsenglish}{%
  \renewcommand{\refname}{Notes and references}
}
\usepackage{array}
\usepackage{droidsans}
\usepackage{charter}
\usepackage{subcaption}
\usepackage[T1]{fontenc}
\usepackage[usenames,dvipsnames]{xcolor}
\usepackage{setspace}
\usepackage[compact]{titlesec}
\usepackage{hyperref}
\usepackage{epstopdf}%This line makes .eps figures into .pdf - please comment out if not required.
\definesubmol\Am[H_2N]{NH_2}

\definecolor{cream}{RGB}{222,217,201}

\DeclareUnicodeCharacter{2212}{\textminus}
\begin{document}

\pagestyle{fancy}
\thispagestyle{plain}
\fancypagestyle{plain}{
%%%HEADER%%%
\renewcommand{\headrulewidth}{0pt}
}
%%%END OF HEADER%%%

%%%PAGE SETUP - Please do not change any commands within this section%%%
\makeFNbottom
\makeatletter
\renewcommand\LARGE{\@setfontsize\LARGE{15pt}{17}}
\renewcommand\Large{\@setfontsize\Large{12pt}{14}}
\renewcommand\large{\@setfontsize\large{10pt}{12}}
\renewcommand\footnotesize{\@setfontsize\footnotesize{7pt}{10}}
\makeatother

\renewcommand{\thefootnote}{\fnsymbol{footnote}}
\renewcommand\footnoterule{\vspace*{1pt}% 
\color{cream}\hrule width 3.5in height 0.4pt \color{black}\vspace*{5pt}} 
\setcounter{secnumdepth}{5}

\makeatletter 
\renewcommand\@biblabel[1]{#1}            
\renewcommand\@makefntext[1]% 
{\noindent\makebox[0pt][r]{\@thefnmark\,}#1}
\makeatother 
\renewcommand{\figurename}{\small{Fig.}~}
\sectionfont{\sffamily\Large}
\subsectionfont{\normalsize}
\subsubsectionfont{\bf}
\setstretch{1.125} %In particular, please do not alter this line.
\setlength{\skip\footins}{0.8cm}
\setlength{\footnotesep}{0.25cm}
\setlength{\jot}{10pt}
\titlespacing*{\section}{0pt}{4pt}{4pt}
\titlespacing*{\subsection}{0pt}{15pt}{1pt}
%%%END OF PAGE SETUP%%%

%%%FOOTER%%%
\fancyfoot{}
\fancyfoot[LO,RE]{\vspace{-7.1pt}\includegraphics[height=9pt]{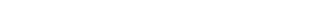}}
\fancyfoot[CO]{\vspace{-7.1pt}\hspace{13.2cm}\includegraphics{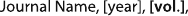}}
\fancyfoot[CE]{\vspace{-7.2pt}\hspace{-14.2cm}\includegraphics{head_foot/RF}}
\fancyfoot[RO]{\footnotesize{\sffamily{1--\pageref{LastPage} ~\textbar  \hspace{2pt}\thepage}}}
\fancyfoot[LE]{\footnotesize{\sffamily{\thepage~\textbar\hspace{3.45cm} 1--\pageref{LastPage}}}}
\fancyhead{}
\renewcommand{\headrulewidth}{0pt} 
\renewcommand{\footrulewidth}{0pt}
\setlength{\arrayrulewidth}{1pt}
\setlength{\columnsep}{6.5mm}
\setlength\bibsep{1pt}
%%%END OF FOOTER%%%

%%%FIGURE SETUP - please do not change any commands within this section%%%
\makeatletter 
\newlength{\figrulesep} 
\setlength{\figrulesep}{0.5\textfloatsep} 

\newcommand{\topfigrule}{\vspace*{-1pt}% 
\noindent{\color{cream}\rule[-\figrulesep]{\columnwidth}{1.5pt}} }

\newcommand{\botfigrule}{\vspace*{-2pt}% 
\noindent{\color{cream}\rule[\figrulesep]{\columnwidth}{1.5pt}} }

\newcommand{\dblfigrule}{\vspace*{-1pt}% 
\noindent{\color{cream}\rule[-\figrulesep]{\textwidth}{1.5pt}} }

\makeatother
%%%END OF FIGURE SETUP%%%

%%%TITLE, AUTHORS AND ABSTRACT%%%
\twocolumn[
  \begin{@twocolumnfalse}
{\includegraphics[height=35pt]{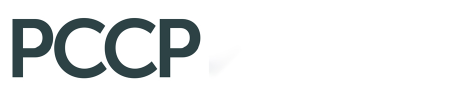}\hfill\raisebox{0pt}[0pt][0pt]{\includegraphics[height=55pt]{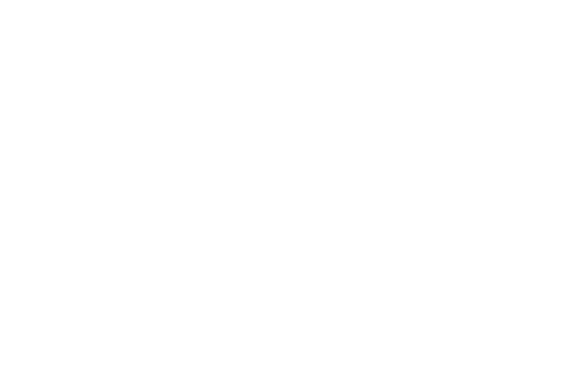}}\\[1ex]
\includegraphics[width=18.5cm]{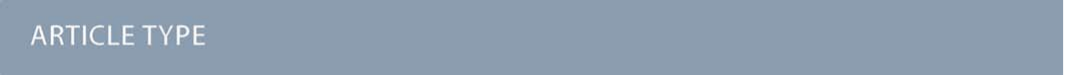}}\par
\vspace{1em}
\sffamily
\begin{tabular}{m{4.5cm} p{13.5cm} }

%%%%%%%%%%%%%%%%%%%%%%%% TITLE %%%%%%%%%%%%%%%%%%%%%%%%

\includegraphics{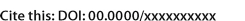} & \noindent\LARGE{\textbf{Carbon Nitride Monolayer Nanosheets: Astrochemical Insights into the Fate of Interstellar Hydrogen}}
%%% Carbon Nitride Monolayer Nanosheets: Insights into the Fate of Interstellar Hydrogen and Astrochemical Considerations
\\%Article title goes here instead of the text "This is the title"
\vspace{0.3cm} & \vspace{0.3cm} \\

 & \noindent\large{David Dubois,\textit{$^{\dag}$}\textit{$^{a}$}\textit{$^{\ddag}$} Pierre Guichard,\textit{$^{\dag}$}\textit{$^{b}$} and Rémi Pasquier\textit{$^{c}$}} \\

%%%%%%%%%%%%%%%%%%%%%% ABSTRACT %%%%%%%%%%%%%%%%%%%%%%

\includegraphics{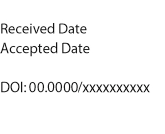} & \noindent\normalsize{Ubiquitously found in the Universe, atomic hydrogen represents up to $70\%$ of the neutral gas composition of the Milky Way. As an adatom, hydrogen can physisorb or chemisorb onto interstellar dust grains and icy mantles, thereby contributing to the formation of \ce{H2} and, potentially, to the synthesis of more complex hydrogenated species. In addition, structures of relatively large specific surface areas --such as silicates, amorphous carbon, graphene sheets, or water ice-- host heterogeneous chemistry that is thought to facilitate the emergence of complex organic matter in astrophysical environments. Although the fundamental physical and chemical processes occurring at dust/gas interfaces are well characterized, current understanding of dust properties governing the formation of \ce{H2} and complex molecules remains incomplete. In this context, we introduce graphitic-like two-dimensional carbon nitride monolayer structures (2D-CN) as a putative molecular family of potential relevance to astrochemistry. The physicochemical and electronic properties of these materials have been extensively examined in recent years for industrial and technological applications. Here, we propose that their importance may likewise extend to interstellar and circumstellar environments. To explore this possibility, we employed Density Functional Theory (DFT) calculations to investigate the characteristics and extent of H adsorption onto \ce{C2N1}, \ce{C3N1}, \ce{C3N2}, \ce{C3N4}, \ce{C4N3}, \ce{C6N6}, \ce{C6N8}, \ce{C9N4}, and \ce{C9N7} monolayer nanosheets. We identify multiple adsorption sites over \ce{C-C} bonds, above C and N atoms, and hollow (macropore) locations at which energetically favorable binding of atomic hydrogen could occur in the interstellar medium (ISM). From an astrochemical perspective, these 2D-CN structures, if formed, could therefore contribute to the physicochemical processing and evolution of hydrogen in the ISM. As such, given their structural similarities to prebiotic nitrogen-bearing frameworks (many found in meteoritic samples and organic aerosols), 2D-CN molecules may emerge as promising candidates for exploring the complex interstellar chemistry of astrophysically-relevant molecules.} \\
\end{tabular}

 \end{@twocolumnfalse} \vspace{0.6cm}

]
%%%END OF TITLE, AUTHORS AND ABSTRACT%%%

%%%FONT SETUP - please do not change any commands within this section
\renewcommand*\rmdefault{bch}\normalfont\upshape
\rmfamily
\section*{}
\vspace{-1cm}

%%%FOOTNOTES%%%

\footnotetext{\dag~These authors contributed equally to this work}
\footnotetext{\textit{$^{a}$~NASA Ames Research Center, Space Science \& Astrobiology Division, Astrophysics Branch, Moffett Field, CA, USA}}
\footnotetext{\textit{$^{b}$~Université de Strasbourg, CNRS, Institut de Physique et Chimie des Matériaux de Strasbourg, UMR 7504, 67000 Strasbourg, France}}
\footnotetext{\textit{$^{c}$~Institute of Theoretical Physics and Regensburg Center for Ultrafast Nanoscopy (RUN), University of Regensburg, 93053 Regensburg, Germany}}

%Please use \dag to cite the ESI in the main text of the article.
%If you article does not have ESI please remove the the \dag symbol from the title and the footnotetext below.

% \footnotetext{\dag~Supplementary Information available: [details of any supplementary information available should be included here]. See DOI: 00.0000/00000000.}
%additional addresses can be cited as above using the lower-case letters, c, d, e... If all authors are from the same address, no letter is required
% \ddag
\footnotetext{\ddag~Present address: Centre de Recherche sur les Ions, les Matériaux et la Photonique CIMAP Normandie Univ, ENSICAEN, UNICAEN, CEA, CNRS, Caen, 14000, France}

%%%END OF FOOTNOTES%%%

%%%MAIN TEXT%%%%
% The main text of the article\cite{Mena2000} should appear here.

% \subsection{This is the subsection heading style}
% Section headings can be typeset with and without numbers.\cite{Abernethy2003}

% \subsubsection{This is the subsubsection style.~~} These headings should end in a full point.  

% \paragraph{This is the next level heading.~~} For this level please use \texttt{\textbackslash paragraph}. These headings should also end in a full point.

\section*{Introduction}
\label{Section 1}

Macromolecular structures are usually separated in four categories: proteins, nucleic acids, polysaccharides, and lipids, all exhibiting specific and identifiable atomic arrangements which result in unique biological, chemical, and physical properties. Due to their nature consisting in the polymerization of smaller subcell units called monomers, these structures allow for many different structural geometries, functionalities (self-organizing, assembly), and sizes which can reach up to several hundreds of thousands of Daltons. Identifying structures of similar mass or complexity in the solar system and the ISM has proven to be sparse, although some proto-organics (\textit{i.e.}, prebiotic molecules) have been found in astrophysical environments \cite{Rehder2010, parker2017formation, Herrero2022, McGuire2022} and meteoritic materials alike \cite{Materese2013, sillerud2024prebiotic}. Over the past decades, the theoretical and experimental investigations of dust grains in the ISM have seen significant advancements, demonstrating their crucial catalytic role for surface chemistry, facilitating the formation of \ce{H2}, and even certain biologically-relevant molecules \cite{Schneiker2022}. On the surface of large dust grains, many mechanisms have been proposed to explain surface alteration which involve adsorption, ion-ion collisions and scattering, water clustering, or diffusion processes \cite{Cui2024}. Under high-energy radiation, the surface of these icy grains undergo important ion$\rightarrow$molecule$\rightarrow$grain reactions that ultimately modify the chemical, morphological, and optical properties of interstellar grains. The Eley-Rideal mechanism, for example, involves the adsorption of a gas phase molecule onto a grain which initiates a surface reaction, followed by the desorption of the product \cite{Cui2024}. Such reactions can occur on silicate grains or amorphous carbon material, and may allow for the formation of a weak hydrogen-bonded catalytic system \cite{fraser2002laboratory}. The Langmuir-Hinshelwood mechanism is another pathway leading to interstellar chemical growth. Here, thermalized adsorbed molecules stochastically diffuse until they reach a binding site, before being lost to form a new product \cite{arumainayagam2019extraterrestrial,Cui2024}. In any case, the potential energy surface (PES) that regulates the physico-dynamics of atoms and molecules on surfaces highlights two pathways, chemisorption and physisorption. Physisorption is primarily barrierless in the cold interstellar environments and as a consequence the adsorbate binding energy is equal to the activation energy for desorption \cite{Potapov2021}. Therefore, identifying sites of potential interest for atomic and molecular adsorption that can also be relevant to astrophysical environments remains fundamental.

In the ISM, polycyclic aromatic hydrocarbons (PAHs) are thought to contain up to $\sim30\%$ of the total interstellar carbon content \cite{tielens2008interstellar, Herrero2022}, while pure carbon allotropes may contain up to $\sim60\%$ of all interstellar carbon matter according to some estimates \cite{mishra2015probing}. The exact composition of the interstellar carbonaceous reservoir is far from being fully characterized, but recent discoveries have unveiled C$_{60}$ and C$_{60}^+$ \cite{berne2012formation, berne2017detection}, nitrogen-containing PAHs (NPAHs) \cite{parker2017formation, wang2024interstellar, Kroonblawd2019}, and graphene \cite{Chen2017, Li2019, sivaraman2023n} as new carbon- and nitrogen-rich candidates relevant to planetary and astrophysical environments. Heterocyclic nitrogenated molecules (cycles having at least one of the carbon inside the ring substituted for a nitrogen atom) remain poorly characterized, although they are likely to play an important role in the formation of NPAHs \cite{Etim2021, Heitkmper2022} and proto-organic molecules such as proteinogenic amino acids \cite{parker2017formation, Kroonblawd2019, wang2024interstellar, Karri2025}. Of these NPAHs, none have been successfully detected yet, although formation pathways of pyrimidine (\ce{C4H4N2}, where two nitrogen atoms have been incorporated into its cycle) and its role in nucleobase synthesis have been investigated \cite{Materese2013, Majumdar2015, nuevo2014photochemistry}. Searches for other heterocyclic aromatic molecules in the ISM such as pyridine, quinoline, or imidazole have so far been unsuccessful \cite{charnley2005astronomical, Etim2021, McGuire2022}. Several factors may explain these non-detections. First, these compounds are expected to be present in low abundance as opposed to PAHs and other organics. They are also expected to photodissociate more rapidly than hydrocarbon chains and PAHs under the combined stellar and interstellar ultraviolet (UV) fields found in protoplanetary nebulae or evolved carbon-rich star envelopes, although these destruction rates are highly dependent on the ring composition of these N-heterocycles \cite{charnley2005astronomical, peeters2005formation}. Second, their emission lines are weak which results from large partition functions and small rotational constants. With many energy levels easily populated, the population of any single level is thus relatively low, which renders their spectral line identification difficult. Nonetheless, searching for these molecules in dense molecular clouds or warmer inner regions of circumstellar envelopes (CSE) may be fruitful in future observation surveys \cite{charnley2005astronomical}, as supported by 6.2 $\mu$m observations attributed to NPAHs in starburst-dominated galaxies \cite{canelo2018variations}. In addition, a number of experimental and theoretical studies have demonstrated that NPAH formation in interstellar-like ices is facilitated even under low-temperature ($10<T<200$ K) conditions relevant to molecular clouds and CSEs \cite{peeters2005formation, sun2014theoretical, parker2015gas, Pentsak2024}, cometary and meteoritic materials \cite{callahan2011carbonaceous, menor2013new, materese2015n, Matuszewski2025}, and planetary atmospheres \cite{Minard1998, ricca2001computational, Gautier2014} where triazine-based ring units have been observed in laboratory-produced aerosols \cite{Derenne2012}. The discovery of naphthalene in cyano-substituted PAH-containing regions has also opened motivation to search for PAHs with nitrogen incorporated into their structures \cite{vats2023theoretical}. Although no N-containing PAHs have been directly detected in the ISM thus far, their detection in meteorites indicates that barrierless or low-barrier pathways leading to their formation are possible under UV-induced photochemistry. Theoretical \ce{C2H2}, HCN, and ion/radical-molecule-based pathways are still up for examination \cite{Pentsak2024}. The importance of these molecules and their formation pathways is indeed high since they are one step away from molecules of prebiotic interest such as nucleobases. Under more extreme (high pressure) conditions, acetonitrile polymerization has been found to occur and is initiated by hydrogen migration reactions which can lead to the formation of graphitic polymers \cite{zheng2016polymerization}. More recently, N-graphene materials were found to be produced after subjecting benzonitrile (\ce{C7H5N}) ice to low-energy photons ($<9$ eV) under vacuum conditions \cite{sivaraman2023n}. These laboratory experiments demonstrated that the formation of N-doped compounds may be possible under low-energy irradiation, thus pointing to the existence of previously unknown bottom-up pathways leading to more complex (and nitrogen-enriched) structures forming under astrochemically-relevant conditions. A recent study by Li et al. \cite{li2026high} combined mass spectrometry measurements with quantum chemistry calculations to investigate potential pathways leading to the formation of carbon nitride structures in an astrophysical context from the reactive \ce{C2N3H} precursor. This study found that its electronic character would promote polymerization from linear to more complex cyclic and fused-ring clusters in the interstellar medium.

Since abiogenic processes are, at present, the only tenable explanation for the presence of nucleobases found in extraterrestrial material, N-containing heterocycles are promising intermediate candidates due to their close resemblance to nucleobases. The five nucleobases found in the deoxyribonucleic acid (DNA) and ribonucleic acid (RNA) backbones (cytosine, guanine, adenine, thymine, and uracil), which participate in the storage of genetic information, share similar structures to pyrimidine (\ce{C4H4N2}) and purine (\ce{C5H4N4}), molecules of astrophysical interest, with two and four nitrogen atoms incorporated into their ring(s), respectively. As described above, these N-heterocycles are sensitive to UV photon absorption. Thus, UV-induced photochemistry plays a crucial role in destruction mechanisms leading to the formation of nucleobases as those found in meteorites \cite{sun2014theoretical, materese2015n, materese2018photochemistry, sandford2020prebiotic}, highlighting completely abiogenic pathways for their production. Furthermore, N-heterocycles containing at least three N atoms such as 1,3,5-triazine (\ce{C3H3N3}) have also been studied \cite{peeters2005formation}. This molecule, perfectly symmetric and with no permanent dipole moment (thus being excluded from any possible rotational spectroscopic search), contains an alternating \ce{C=N} covalent bond sequence. Photolysis experiments of 1,3,5-triazine result in the formation of the :\ce{C#N-C#N} radical \cite{peeters2005formation}, a backbone thought to be relevant to the atmosphere of Saturn's largest moon Titan and cold molecular clouds \cite{Bierbaum2011a, Dubois2019b}. The lifetime of 1,3,5-triazine and pyrimidine is expected to be very short on astronomical timescales in the solar system and diffuse ISM (from minutes to several years), but relatively large in dense molecular clouds (up to 1 Myr) \cite{peeters2005formation}. This would allow for their survival until the end of life of these clouds.
%%% check paper by Pascal on tholinomics on NPAHs...

% \chemfig{C*6(-N=C(-NH_2)-N=C(-NH_2)-N(-NH_2)=)}

In this context, we propose in this paper a new possible class of astrochemically-relevant polymeric structure, namely graphitic-like two-dimensional carbon nitrides (hereafter called 2D-CN) which, we suggest, may be of particular relevance to nitrogen-rich environments in the interstellar medium. In the present study, we focus on the role of hydrogen adsorption onto 2D-CNs. Because of their unique molecular and electronic properties, these large molecular structures exhibit a unique opportunity for the astrochemistry community to explore their spectroscopic and (photo)chemical properties in the future. In addition, their potential for atomic and molecular adsorption warrants further scrutiny into their applications for astrochemistry. We briefly overview 2D-CN nanosheets and their properties and applications. We then describe our computational method using DFT which we employed to conduct a survey of H adsorption onto \ce{C2N1}, \ce{C3N1}, \ce{C3N2}, \ce{C3N4}, \ce{C4N3}, \ce{C6N6}, \ce{C6N8}, \ce{C9N4}, and \ce{C9N7} frameworks. Finally, we discuss potential astrophysical implications on the fate and evolution of interstellar atomic hydrogen, its adsorptive character on carbon nitrides, and heterogeneous chemical complexity.\\

\section*{2D Carbon Nitride Nanosheets: Structures, Applications, and Properties}
\label{Section 2}
% \subsection{Industrial applications}

\begin{figure*}
    \centering
    \includegraphics[width=1.0\textwidth]{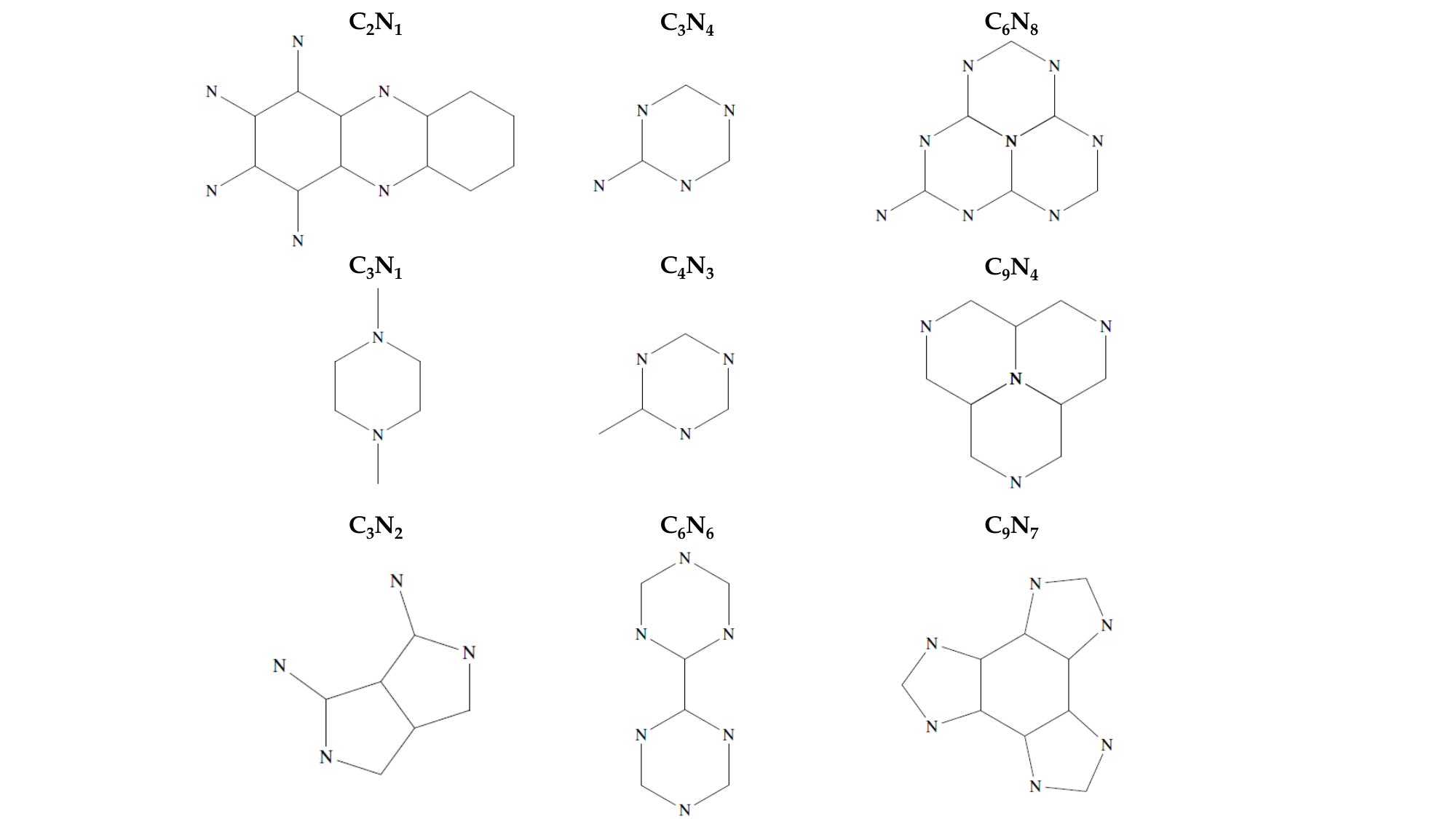}
    \caption{Schematic drawings of the unit cells for each of the studied structures.}
    \label{fig:1}
\end{figure*}

Graphitic carbon nitride–based nanosheets have attracted considerable attention over the past three decades owing to their distinctive electronic, magnetic, and physicochemical characteristics, which make them highly relevant for industrial and nanomaterial applications (the reader is referred to the comprehensive reviews by Zhao et al. \cite{Zhao2014} and Rono et al. \cite{Rono2021Review}). For example, \ce{C3N2} and \ce{C4N} display metallic and semimetallic behavior, respectively, whereas \ce{C2N1}, \ce{C3N1}, \ce{C3N4}, \ce{C6N6}, and \ce{C6N8} all exhibit semiconducting properties. Consequently, these two-dimensional carbon nitride frameworks have played a significant role in advancing technologies in energy generation and conversion, metal-free catalysis, and spintronics, as evidenced by numerous theoretical and experimental studies \cite{Wang1997,Li2013PCCP,AlgaraSiller2014,Hashmi2014,Hu2014SR,Zhao2014,Zhu2014,Hashmi2015,Liu2015,Bafekry2019,Rao2019,Mehrdad2020,Subhan2020,Bhowmick2021,Liu2021,Oseghe2021,Rono2021JMSL,Rono2021Review,Muchuweni2022,Rezapour2022,Lal2024,Naseem2024,Rono2025,Yuan2025}. While graphene displays carbons in an $sp^2$ configuration, carbon nitrides such as \ce{C3N4} contain $\pi$-conjugated sheets via the $sp^2$ hybridization of C and N atoms, giving it its unique electronic structure where the lone pair from N is stabilized by the $\pi$ bonding \cite{Rono2021Review}. Schematic drawings of the unit cells for each of the studied structures are shown in Figure \ref{fig:1}. 

% \begin{figure}[h]
%     \centering
%     % \includegraphics[width=0.5\linewidth]{}
%     \chemfig{*6((-!\Am)-N=(-!\Am)-N=(-!\Am)-N=)}
%     \caption{Molecular structure of the s-triazine (\ce{C3N3}) skeleton, part of the most common derivative melamine \ce{C3H6N6} shown as an example.}
%     \label{fig:1}
% \end{figure}

% \begin{figure*}
%     \centering
    
% \textbf{\ce{C2N1}} \hspace{7cm} \textbf{\ce{C3N4}} \hspace{4cm} \textbf{\ce{C6N8}} \vspace{0.2cm}

%     \chemfig{*6((-)-N-*6(-----)--N-*6(-(-N)-(-N)-(-N)-(-N))-)} \hspace{1.2cm} \chemfig{*6((-N)-N--N--N-)} \hspace{3cm} \chemfig{
%         *6((-N)
%             -N-
%             *6(
%                 -N-
%                 -N-
%                 *6(
%                     -N-
%                     -N-
%                     -N-
%                     -
%                 )
%                 -N
%                 -
%             )
%             -N-
%             -N-
%             -
%         )
%     }

% \vspace{0.5cm}

% \textbf{\ce{C3N1}} \hspace{7cm} \textbf{\ce{C4N3}} \hspace{4cm} \textbf{\ce{C9N4}} \vspace{0.2cm}
    
%     \chemfig{*6(-N(-)---N(-)---)} \hspace{5cm} \chemfig{*6((-)-N--N--N-)} \hspace{4cm} \chemfig{
%         *6(
%             -N--
%             *6(
%                 --N
%                 --
%                 *6(
%                     --N
%                     --
%                     -N-
%                     -
%                 )
%                 -N
%                 -
%             )
%             -N-
%             --
%             -
%         )
%     }

% \vspace{0.5cm}

% \textbf{\ce{C3N2}} \hspace{7cm} \textbf{\ce{C6N6}} \hspace{4cm} \textbf{\ce{C9N7}} \vspace{0.2cm}

% \chemfig{N*5(--*5(--N-(-N)-)--(-N)--)} \hspace{3cm} \chemfig{*6(-N--N-*6(-*6(-N--N--N-))-N--N)} \hspace{3cm} \chemfig{*6(-*5(-N--N-)--*5(-N--N-)--*5(-N--N-)-)}
    
%     \caption{Schematic drawings of the unit cells for each of the studied structures.}
%     \label{fig:1}
% \end{figure*}

As a bulk material, \ce{C3N4} possesses a band gap of 2.7 eV and a low specific surface area \cite{Rono2021Review}. In contrast to its bulk characteristics, the photocatalytic performance of \ce{C3N4} is significantly enhanced after its exfoliation into nanosheets or quantum dots \cite{Rono2021Review}. For a comprehensive overview of the available exfoliation mechanisms to produce carbon nitride nanosheets, the reader is referred to the review by Rono et al. \cite{Rono2021Review}. Other structures of strong covalency and where nitrogen is more uniformly distributed (\textit{e.g.}, in the case of \ce{C3N1}) result in high thermal conductivity where the electronic properties can be tuned according to the presence of adatoms and admolecules \cite{Bafekry2019}. Other nanosheets (\textit{e.g.}, \ce{C6N6}, \ce{C6N8}, \ce{C9N4}) display larger surface areas due to porous surfaces, making them important surfaces for catalytic reactions \cite{hsu2024graphene,Lal2024,Naseem2024}. In particular, large ($\sim5$~\AA) pores such as those found in \ce{C2N1} or \ce{C9N4} (see below) are circled by six N pyridinic $sp^2$ hybridized atoms, thus forming an electron-rich cavity. Therefore, this configuration creates efficient reactive sites which can attract electron-poor species and improve their catalytic behavior \cite{Faisal2017}. The investigation of 2D-CNs therefore continues to be a dynamic and rapidly evolving field of research \cite{Rono2021Review, Wang2022}.\\

\section*{Methods}
\label{Section 3}

The adsorption of hydrogen on the surfaces was studied with Density Functional Theory (DFT) calculations using the Quantum ESPRESSO suite \cite{giannozzi2017advanced}. %All simulations employed the Self-Consistent Field (SCF) method to determine the electronic ground state of the two-dimensional carbon–nitride (2D-CN) monolayers with the hydrogen atom investigated in this work. 
The set of studied structures comprises the following compounds: $\mathrm{C}_2\mathrm{N}_1$, $\mathrm{C}_3\mathrm{N}_1$, $\mathrm{C}_3\mathrm{N}_2$, $\mathrm{C}_3\mathrm{N}_4$, $\mathrm{C}_4\mathrm{N}_3$, $\mathrm{C}_6\mathrm{N}_6$, $\mathrm{C}_6\mathrm{N}_8$, $\mathrm{C}_9\mathrm{N}_4$, and $\mathrm{C}_9\mathrm{N}_7$. For each material, the initial geometry was based on previously reported optimized structures available in the literature~\cite{Bafekry2019}. The structures were then relaxed using the VASP program (as we faced convergence difficulties when carrying the optimization in Quantum ESPRESSO)~\cite{Kresse1996}. We employed the Perdew–Burke–Ernzerhof (PBE) functional within the Generalized Gradient Approximation (GGA)~\cite{Perdew1997}. Core-level electrons are treated at the projector-augmented wave (PAW)~\cite{Blochl1994} level. The plane wave cutoff and the k-grid were independently converged using a total energy cutoff of 1~meV/atom for each system. We then carried a constant cell-shape optimization of the system and the unit cell, increasing the previously obtained cutoffs and k-grid size in order to avoid Pulay stress.
% Representative supercells used in these simulations are shown in Fig.~\ref{fig:supercell-gC4N3}.

For the adsorption calculation, we employed the Perdew–Burke–Ernzerhof (PBE) functional within the Generalized Gradient Approximation (GGA)~\cite{Perdew1997}. Core–valence interactions were treated using PAW pseudopotentials from the PSLibrary~\cite{dalcorso2014}: H (1.0079) -- \texttt{H.pbe-kjpaw\_psl.1.0.0.UPF}, C (12.011) -- \texttt{C.pbe-n-kjpaw\_psl.1.0.0.UPF}, N (14.0067) -- \texttt{N.pbe-n-kjpaw\_psl.1.0.0.UPF}. Plane-wave kinetic-energy and charge-density cutoffs were selected to ensure convergence of total energies. The self-consistent (SCF) cycle was converged using an energy threshold of $10^{-9}$ Ry. Brillouin-zone integrations used a $6\times 6\times 1$ Monkhorst-Pack~\cite{monkhorstpack1976} mesh, appropriate for the two-dimensional nature of the systems. A vacuum region exceeding 15 \AA~was included along the out-of-plane direction to avoid spurious interactions between periodic images. The dispersion corrections have been computed using the DFT-D3 scheme~\cite{Grimme2010}, however their contribution is very small and as such they will be neglected in the following (see the Results section and the Supplementary material).

To characterize the interaction between a hydrogen atom and the surfaces of the 2D-CN monolayers, we constructed three-dimensional potential-energy landscapes by systematically sampling the position of a single H atom relative to each monolayer.

For each material, a set of in-plane sampling points was defined across the primitive cell (or supercell), spanning all symmetry-inequivalent regions such as C sites, N sites, bridge sites, and hollow sites. At each fixed lateral coordinate $(x,y)$, the hydrogen atom was displaced along the out-of-plane direction ($z$) from a non-interacting height toward the surface in steps of $\Delta z$. Single-point total-energy calculations were performed at each sampled position, while keeping the monolayer atoms frozen at their optimized geometry. The hydrogen atom was constrained from relaxing laterally in order to obtain consistent sampling of the interaction potential.

From the full three-dimensional sampling grids, one-dimensional potential-energy curves $E_{\mathrm{int}}(z)$ were constructed for each relevant site by fixing $(x,y)$ and varying $z$. The equilibrium adsorption height $z_{\mathrm{eq}}$ was obtained from the minimum of these curves. The adsorption energy at each site was then computed as $E_\mathrm{ads}=E_\mathrm{int}(x_\mathrm{site},y_\mathrm{site},z_\mathrm{ads})$ with negative values indicating exothermic adsorption. Comparison of the minimum values among the different sites allowed us to determine the most favorable adsorption positions for hydrogen on each 2D-CN monolayer. We neglect the zero-point energy contribution to the adsorption properties as these have a negligible impact on the relative energetics of the adsorption sites~\cite{Bai2018}. 

%This samplinbased methodology thus provides a consistent and detailed description of the energetic landscape experienced by a hydrogen atom near various carbon–nitride monolayers, enabling quantitative determination of adsorption energies and preferred binding geometries across a large part of the 2D-CN family.\\

% \begin{figure}[h]
%   \centering
%   \includegraphics[width=1.0\textwidth]{potential-gC4N3.png}
%   \caption{caption}
%   \label{fig:potetial-gC4N3}
% \end{figure}

% \begin{figure*}
%  \centering
%  \includegraphics[height=10cm]{potential-gC4N3.png}
%  \caption{A two-column figure.}
%  \label{fgr:example2col}
% \end{figure*}

%\begin{figure*}
% \centering
% \includegraphics[height=10cm]{C4N3.png}
% \caption{Computational super-cell of $\mathrm{g}-\mathrm{C}_4\mathrm{N}_3$. The dark atoms are carbons and %the blue-greys are the nitrogens.}
% \label{fig:supercell-gC4N3}
%\end{figure*}

\section*{Results}
\label{Section 4}

The results for our calculations are displayed in Figure~\ref{fig:POT} (and the associated DFT-D3 dispersion-correction available in the Supplementary Material). As one can see in the lower part of the figure, the adsorption sites have been sampled on the high-symmetry points of each respective lattice. For the C$_2$N$_1$ system, the adsorption sites were sampled in the middle of the C-N and C-C bonds, along with the the hexagonal and macropore (Centre 3) site and on top of the carbon and nitrogen atoms. The results show that the privileged site is located on top of the nitrogen atom, with a minimum of -1.71~eV. For the case of \ce{C3N4}, the adsorption site is located in-plane within the macropore, with a very deep minimum of -2.92~eV. The case of \ce{C6N8} is interesting, as the system does not show a strongly privileged adsorption site, with a shallow minimum of -0.22~eV above a C atom. Very similar results can be observed for \ce{C3N1}, with a minimum of -0.24~eV at the top of the carbon site. For \ce{C4N3}, we observe two major minima: one on top of the macropore with a very deep value of -5.37~eV (Centre 2) and several secondary minima over the carbon and nitrogen atoms. \ce{C9N4} shows two almost degenerate minima between the bridge C-C site and the C atomic site, with a slightly lower minimum in the second case of -1.06~eV. \ce{C3N2} shows several adsorption minima, with the major one being located in-plane in the NN1 site as defined in Figure~\ref{fig:POT}g with a value of -4.18~eV and several secondary sites located on top of all the other high-symmetry points. \ce{C6N6} shows a major minimum on top of the nitrogen atom, with a value of -1.57~eV. Finally, \ce{C9N7} shows a major minimum on top of the C-C bond in the bridge position, with several minima on the other symmetry points including a deep secondary minimum on top of the carbon atom. Owing to the strong adsorption of the hydrogen atom on most of the surfaces, which is for most systems in the chemisorption regime (above 0.5~eV), the contribution of dispersion correction to the adsorption energetics is in most cases very small (below 0.1~eV, see Figure~S1 in the Supplementary materials). One can also rationalize this by remembering that dispersion corrections are a semi-empirical contribution to take into account Van der Waals forces between the surface and the hydrogen atom, which is weakly polarizable and therefore exhibit weak dispersion effects.  We have to note two counter-examples, being C$_6$N$_6$ and C$_6$N$_8$, where the adsorption is weaker so that physisorption processes become more important, although in that case the dispersion corrections still do not change the qualitative picture such as the location of the privileged adsorption site as one can see from Figures~\ref{fig:POT} and S1.).
A summary of the results is given in Table~\ref{Table 1}. As such, the general adsorption properties are system dependent, and an in-depth understanding of the relative energetics of the adsorption sites would require a careful study of the chemical properties of the surfaces which is beyond the scope of this paper. However, we do observe an adsorption site in each case (although fairly shallow in some cases such as \ce{C3N1}), which indicates that graphitic carbon nitrides are indeed strong candidates for hydrogen adsorption, in accordance with the already established results in the literature~\cite{Panigrahi2020}. Our work being oriented towards gas-phase systems for astrochemical purposes,  bulk-phase properties are out of the scope of this paper; however,  we have to mention that as emphasized and carefully demonstrated by Bafekry~\cite{Bafekry2019}, one can then show that this strong hydrogen adsorption leads to a significant transformation of the physical and chemical properties of the 2D graphitic carbon nitrides, with for example significant charge transfers or changes in the band structure of the systems such as their band gaps or magnetism.

\begin{table*}[h!]
\centering
\caption{Molar mass and atomic composition of selected carbon nitride monolayer super-cells considered here. Main adsorption sites with the potential depth with the DFT-D3 correction are shown in Figure \ref{fig:POT}. \label{Table 1}.}
\begin{tabular}{lcccccccc}
\hline
Monolayer & $\Sigma$ C & $\Sigma$ N & \textit{a} (\AA) & d$_{\text{C-C}}$(\AA) & d$_{\text{C-N}}$(\AA) & Main adsorption site & Potential (eV) & D3 Correction (eV) \\
\hline
\ce{C2N1} & $42$ & $21$ & $8.32$ & $1.43-1.47$ & $1.34$ & N atom & $-1.71$ & $-0.037$\\
\ce{C3N1} & $48$ & $16$ & $4.86$ & $1.40$ & $1.40$ & C atom & $-0.24$ & $-0.064$ \\
\ce{C3N2} & $39$ & $26$ & $8.32$ & $1.36-1.47$ & $1.33-1.36$ & NN1 & $-4.18$ & $-0.039$ \\
\ce{C3N4} & $27$ & $36$ & $4.78$ &  & $1.33-1.46-1.47$ & Center 2 & $-2.92$ & $-0.041$ \\
\ce{C4N3} & $36$ & $27$ & $4.84$ & $1.44$ & $1.35$ & Center 2 & $-5.37$ & $-0.052$ \\
\ce{C6N6} & $36$ & $36$ & $7.12$ & $1.51$ & $1.34$ & N atom & $-1.57$ & $-0.034$\\
\ce{C6N8} & $30$ & $40$ & $7.13$ &  & $1.33-1.39-1.47$ & C atom & $-0.22$ & $-0.038$ \\
\ce{C9N4} & $36$ & $27$ & $9.63$ & $1.42-1.45$ & $1.34-1.39$ & C atom & $-1.06$ & $-0.043$\\
\ce{C9N7} & $36$ & $27$ & $8.05$ & $1.42-1.47$ & $1.33-1.37-1.42$ & C-C bond & $-2.40$ & $-0.023$

\\
\hline
\end{tabular}
\end{table*}

% \ce{C2N1} & $42$ & $21$ & $8.324$ & $1.4282-1.4683$ & $1.3364$ & N atom & $-1.71$\\
% \ce{C3N1} & $48$ & $16$ & $4.859$ & $1.4032$ & $1.4022 $ & C atom & $-0.24$ \\
% \ce{C3N2} & $39$ & $26$ & $8.320$ & $1.3635-1.4680 $ & $1.3282-1.3646 $ & NN1 & $-4.18$ \\
% \ce{C3N4} & $27$ & $36$ & $4.783$ &  & $1.3268-1.4569-1.4723 $ & Center 2 & $-2.92$ \\
% \ce{C4N3} & $36$ & $27$ & $4.836$ & $1.4413 $ & $1.3477 $ & Center 2 & $-5.37$ \\
% \ce{C6N6} & $36$ & $36$ & $7.116$ & $1.5080$ & $1.3400$ & N atom & $-1.57$ \\
% \ce{C6N8} & $30$ & $40$ & $7.132$ &  & $1.3321-1.3925-1.4749$ & C atom & $-0.22$ \\
% \ce{C9N4} & $36$ & $27$ & $9.634$ & $1.4214-1.4547$ & $1.3400-1.3933$ & C atom & $-1.06$ \\
% \ce{C9N7} & $36$ & $27$ & $8.050$ & $1.4189-1.4720$ & $1.3337-1.3690-1.4196$ & C-C bond & $-2.40$ 

\begin{figure}
    \centering
    \includegraphics[width=1\linewidth]{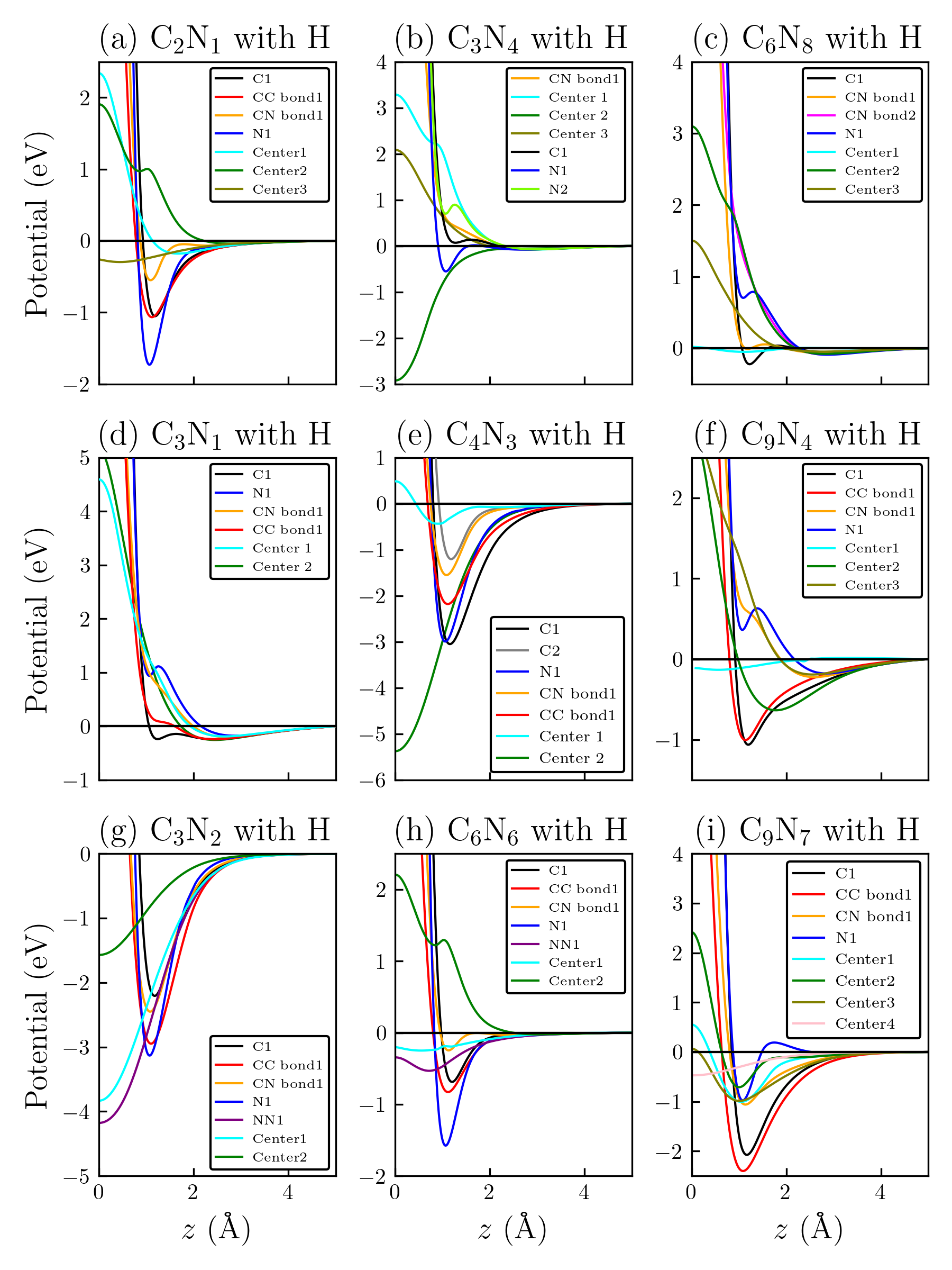}
    \includegraphics[width=1\linewidth]{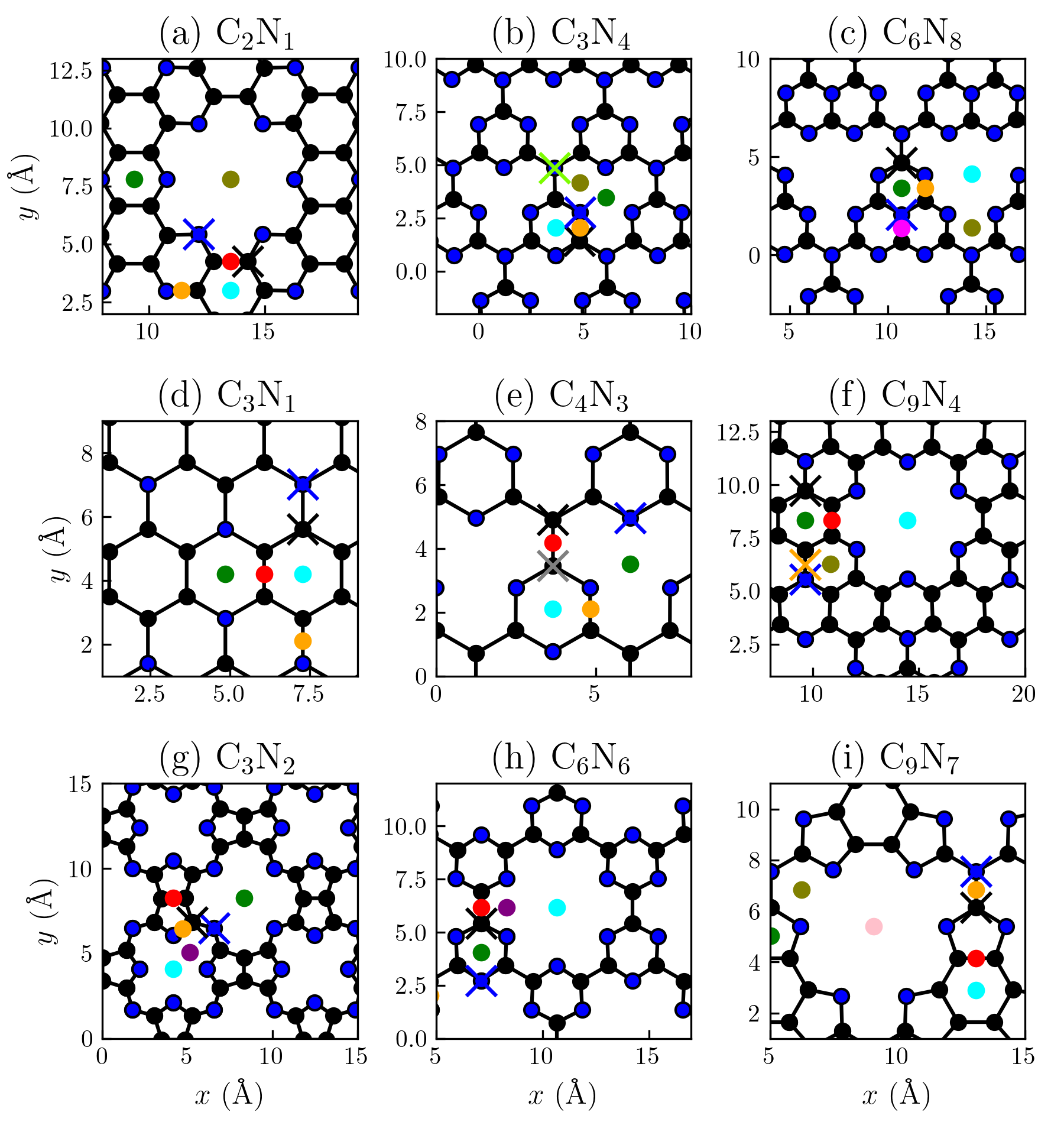}
    \caption{(Upper) Potential energy as a function of the distance from the sheet for a hydrogen atom at a specific sheet position. (Lower) Sheet representation showing the corresponding calculation sites: (a) $\mathrm{C}_2\mathrm{N}_1$, (b) $\mathrm{C}_3\mathrm{N}_4$, (c) $\mathrm{C}_6\mathrm{N}_8$, (d) $\mathrm{C}_3\mathrm{N}_1$, (e) $\mathrm{C}_4\mathrm{N}_3$, (f) $\mathrm{C}_9\mathrm{N}_4$, (g) $\mathrm{C}_3\mathrm{N}_2$, (h) $\mathrm{C}_6\mathrm{N}_6$, (i) $\mathrm{C}_9\mathrm{N}_7$. Black atoms are carbons and the blue atoms are nitrogens. Coloured dots are referring to the specific positions calculated in the sheet.}
    \label{fig:POT}
\end{figure}

\section*{Astrophysical implications}
\label{Section 5}
\subsection*{Interstellar hydrogen chemistry and H$_2$ formation}
\label{section 5.1}

Neutral atomic and molecular hydrogen are found ubiquitously across the interstellar medium. It has generally been accepted that \ce{H2} is primarily formed on dust grains and not in the gas phase \cite{Vidali2013}. Studies of \ce{H2} formation on dust grain analogues (\textit{i.e.}, silicates, amorphous carbon, and organic ices) have answered numerous questions about the surface mechanisms at low temperature ($<20$ K) while raising additional questions concerning kinetic timescales, impact of surface properties, and temperature effects on \ce{H2} formation mechanisms \cite{Vidali2013}. At warmer temperatures (up to $\sim150$ K), the processes leading to \ce{H2} formation remain largely unknown and the variable nature and size of dust grain analogues contributes to this uncertainty \cite{Vidali2013,Bron2014}. The activation energy barriers that H can overcome have been relatively well characterized for graphite. Long range dispersion forces with characteristic energies below $\sim100$ meV correspond to physisorbed conditions, whereas chemisorption involves substantially higher energies on the order of 1 eV to establish chemical bonding. Structural defects and atomic disorder may reduce these energy barriers, a process demonstrated to occur in carbonaceous materials \cite{Vidali2013}.

Though widespread in the ISM, hydrogen is expected to diffuse faster on the surface of grains at warmer temperatures through the Langmuir-Hinshelwood mechanism \cite{Vidali2013}. At even higher temperatures, the Eley-Rideal mechanism is thought to become predominant as H evaporation is more likely to occur in strongly irradiated regions such as the edge of photodissociation regions (PDR). \ce{H2} formation rates in these regions are expected to reach at least $3~\times~10^{-17}~n_\text{H}n(\text{H})$ cm$^3$ s$^{-1}$ (where $n_\text{H}$ represents the proton density in the gas phase (in cm$^{-3}$) and $n(\text{H})$ the number of H atoms that are adsorbed on the surface) \citep{LeBourlot2012,Vidali2013,Bron2014}. It has also been shown that in the high-gas density edge regions of PDRs, rates for Langmuir-Hinshelwood mechanisms can be comparable to those for Eley-Rideal processes \cite{Bron2014}. Other than PDRs, regions where H adsorption plays an important astrochemical role include diffuse and dense clouds, active galactic nuclei, and shocked regions \cite{Vidali2013,Padovani2018}. The question of physisorption and chemisorption processes on (organic) dust materials is thus fundamental for our understanding of heterogeneous growth and \ce{H2} formation, for which few constrained parameters exist. Furthermore, studies of \ce{H2} chemistry in H-graphite and H-coronene systems have shown hydrogen chemisorption to be an efficient process on top of their C-C bonds or over their hollow regions, even down to very cold temperatures \cite{Bonfanti2007,Davidson2014,Barrales-Martnez2018}. This underscores that dihydrogen chemistry involving hydrogen adsorbed on the cold surfaces of interstellar nanoparticles requires further investigations to clarify the fate of interstellar hydrogen.

Our results shows that 2D-CNs can provide a viable pathway to atomic hydrogen chemisorption with the presence of preferential adsorption sites on either \ce{C-C}, above C and N atoms, or macropore sites (Table \ref{Table 1} and Figure \ref{fig:POT}). As such, the proposed Reaction \ref{H2 formation} may take place, as previously suggest in the context of \ce{H2}/graphite systems \cite{Farebrother2000}. Such a reaction is of particular interest since present day cosmic ionization levels do not permit a substantial production of \ce{H2} from the reaction between \ce{H-} and H \cite{Vidali2013}, thus supporting the contribution of heterogeneous chemistry for \ce{H2} formation.

\begin{equation}
    \ce{H + H/\text{2D-CN} \longrightarrow H2 + \text{2D-CN}}
    \label{H2 formation}
\end{equation}

The timescales required for these reactions are difficult to constrain since they are highly substrate composition-dependent and limited experimental constraints (\textit{e.g.}, regarding sticking coefficients or binding energies) exist. Moreover, the composition and atomic structure of ISM dust particles are not fully known known which makes kinetics estimations difficult to constrain. Nonetheless, various laboratory and \textit{ab initio} calculation studies have provided insights into these parameters (see Refs. \cite{Bonfanti2007,Vidali2013} and references therein). One derived parameter, the diffusion coefficient \textit{D} for a surface plane, can be approximated to \cite{Bonfanti2007}:

\begin{equation}
    D = \frac{1}{4}~ a_0^2 ~\Gamma_h
\end{equation}

where $a_0$ is the lattice constant (the distance between adjacent adsorption sites) and $\Gamma_h$ the hopping rate. As an example for \ce{C3N4}, if we approximate the hopping rate to that of a graphitic-like tunneling time of $\Gamma_h=0.9$ ps \cite{Bonfanti2007} and for a lattice constant of 4.78 \AA ~(Table \ref{Table 1}), the diffusion coefficient $D=5.1~\times~10^{-4}$ cm$^2$ s$^{-1}$, compared to $D=1.7~\times~10^{-4}$ cm$^2$ s$^{-1}$ in the case of graphite \cite{Bonfanti2007}. Although obtained from extrapolations of graphitic-like systems, this slightly higher estimated value of $D_{H/2DCN}$ compared to $D_{H/graphite}$ would result from a larger lattice constant for 2D-CN than that of graphite. These are only approximate estimates and future tailored investigations would be needed to constrain the chemistry and dynamics of hydrogen onto these nanosheets. Regarding 2D-CN, the diffusion of atomic hydrogen is governed by both either quantum tunneling or classical thermal diffusion, depending on the surface temperature and the diffusion barrier. At very low temperatures (\( T < 50 \, \text{K} \)), quantum tunneling dominates, allowing H atoms to diffuse rapidly with a diffusion coefficient of approximately \( 6.0 \times 10^{-4} \, ~\text{cm}^2/\text{s} \) and a tunneling time of 0.9 ps, thus permitting H atoms to be very mobile \cite{Casolo2009}. 
% This fast diffusion is due to the light mass of H and the narrow potential barriers on the \ce{C3N4} surface. 
As the temperature increases, classical diffusion becomes more significant. For a physisorption-like barrier of $\approx$ 0.1 eV, the classical diffusion coefficient at 300 K is about \( 1.1 \times 10^{-5} \, \text{cm}^2/\text{s} \), which is slower than tunneling but still allows for some mobility. However, for barriers relevant to chemisorption $\approx$ 0.5 eV, classical diffusion at 300 K is extremely slow (\( D \approx 6.5 \times 10^{-13} \, \text{cm}^2/\text{s} \)), rendering the mobility of H atoms almost null. This stark contrast highlights the importance of the barrier heights: low barriers enable efficient diffusion even at moderate temperatures (Eley-Rideal mechanisms from physisorbed molecules can be efficient even at low temperature \cite{Bonfanti2007}), while high barriers suppress diffusion unless the temperature is significantly elevated (\textit{e.g.}, T $>$ 300 K). In these higher temperature regions (\textit{e.g.}, PDRs, shock regions), hydrogen chemisorption is very favorable and may lead to molecular recombination to form \ce{H2}. In cold astrophysical environments, however, tunneling could in theory facilitate \ce{H2} formation on 2D-CN surfaces, while classical diffusion may only play a role in warmer, diffuse environments or on surfaces with lower energy barriers.

Casolo et al. \cite{Casolo2009} proposed that, given the narrow parameter space (\textit{e.g.}, $n_{(H)}$, temperature) needed for molecular hydrogen formation on a model of pristine graphite in interstellar clouds, hydrogen would have to be primarily physisorbed to enable efficient Eley-Rideal mechanism to operate. Another scenario, they speculate, might be that pristine graphite may be an incomplete model to fully explain hydrogen formation in interstellar cloud conditions. Thus, 2D-CN nanosheets may provide a structural and compositional alternative to such models for future studies. Finally, interstellar hydrogen adatoms would also impact the electronic, structural, and magnetic properties of 2D-CN, as shown by Bafekry et al. \cite{Bafekry2019}. For example, it was found that an H adatom above a C site would, post relaxation, elongate \ce{C-C} and \ce{C-N} bond lengths for certain structures (\textit{e.g.}, \ce{C3N1} and \ce{C3N2}), while for other structures (\textit{e.g.}, \ce{C3N4}, \ce{C6N6}) shift their semiconducting properties to pure metals. For \ce{H/C4N3} adsorption, \ce{C4N3} is seen to transition from a half-metal to a direct semiconductor. Interestingly, in the same study, the impact of atomic vacancies or substitutions of N by H resulted in variations of \ce{C-H} bond lengths and electronic properties of 2D-CN. These observations further highlight the need to explore new chemical properties emerging from H adsorption onto 2D-CN and potential pathways of more complex chemistry thereof.

% As such, our analysis reveals adsorption sites on all 9 2DCN nanosheets studied. The structures with the most favorable adsorption sites are \ce{C2N1}, \ce{C3N1}, \ce{C3N2}, \ce{C3N4}, \ce{C4N3}, \ce{C6N6}, \ce{C9N4}, and \ce{C9N7}.
% Given  2DCN nanosheets 
% In cold ISM regions ($<50$ K), a chemisorbed hydrogen atom is not able to chemically react since at these low tempera
% talk about bond hybridization too

% \begin{equation}
%     \tau_h^{-1} = \nu \exp\left(-\frac{E_{\text{diff}}}{k_B T_s}\right)
% \end{equation}

% with $\nu$ the attempt frequency typically $10^{12}$ s$^{-1}$, E$_{\text{diff}}$ the diffusion barrier, $k_B$ the Boltzmann constant, and $T_s$ the surface temperature.

\subsection*{Carbon nitride nanosheets as chemical catalysts of complex organics in the interstellar medium?}
\label{section 5.2}

In light of our results and the available literature, we conceptually examine three putative chemical growth mechanisms of heterogeneous chemistry that may involve carbon nitrides in the ISM: (a) chemical erosion and photocatalysis, (b) hydrogenation reactions, and (c) H-ion reactions. First, although chemical erosion (\textit{i.e.}, the chemical removal of C or N atoms from a given monolayer) is not directly linked to our study of H adsorption, experimental work has shown that graphitic mantles coated with \ce{H2O}-based ices would proceed to be eroded under UV \cite{Bergeld2004,Shi2015} and visible \cite{Chakarov2005,Potapov2021} wavelength photons. Ice-coated graphite such as \ce{H2O/\text{graphite}} or \ce{CO/\text{graphite}} under photon-driven radiation $>15$ eV can lead to the formation of \ce{H2}, \ce{CO}, or \ce{NH3} \cite{Duley2000,Chakarov2005,Fillion2012}. These photocatalytic mechanisms are also commensurate with the substrate surface temperature, possible atomic defects, and the structural arrangements of ice layers deposited onto them. Moreover, experimental studies have also demonstrated the possibility for chemical growth leading to formaldehyde formation on pristine a-C grains as low as 10 K, which place protoplanetary and interstellar cloud environments at the center of interest to probe heterogeneous chemical processes \cite{Potapov2021}. At Saturn's moon Titan, a widely studied satellite rich in organics, experimental simulations of its atmospheric chemistry by Sekine et al. \cite{Sekine2008} showed that chemical erosion of organic aerosols is an inefficient process under Titan conditions. Future work however is needed to establish a clear relationship between surface temperature and chemical erosion \cite{Shi2015,Hansen2025}. We also note the possible presence of N-rich and CN skeletons composing Titan aerosol analogues studied by Gautier et al. \cite{Gautier2017} and Quirico et al. \cite{Quirico2008}, respectively. As a planetary prototype, Titan aerosols contain mass fragments that could thus be directly linked to 2D-CNs.
Second, growth by sequential hydrogenation has been shown to be an efficient process on astrophysical ices where E-R pathways could form formaldehyde, methanol, or ammonia \cite{Herbst2005,Tieppo2023}. Subsequently, methanol can be an entry point into the growth of more complex organic molecules \cite{Potapov2021}. Graphitized silicon carbide (a wide-gap semiconductor) has also been proposed experimentally as a potential intermediate in top-down formation routes of PAHs \cite{Merino2014}. Under high temperature ($>900$ K) and hydrogen bombardment, the latter experiments highlighted the importance of H adsorption and H clustering (hyperhydrogenation) leading to the formation of PAHs, benzene rings, and acetylene. The variety of samples in composition (organic or mineral surfaces) and conditions (temperature, gas density) applied to a range of astrophysical conditions therefore warrants future studies focused on examining the role of graphitic-like structures on chemical complexity and growth. While out of the scope of this study, another possible scenario involves gas-phase nitrogenated polycyclic structures in PDRs. In this case, sequential H abstraction reactions from positively charged coronene was found to be more efficient than from its neutral counterpart \cite{Foley2018}, a process also shown to be relevant for anthracene and pyrene surface \cite{Barrales-Martnez2018}.

In comparison to amorphous carbon, the study of infrared and X-ray photoelectron spectroscopy bands for amorphous carbon nitrides has been investigated by several groups \cite{Chowdhury1998,Rodil2002,Wang2008,McMillan2009,Suter2018}. These works have unveiled a prominant $sp^2$ character due to the presence of \ce{-C=C-} and \ce{-C-N=} bonding content. An increase in nitrogen incorporation also induces an increase in intensity of the \ce{N=N}, \ce{C#N}, and \ce{C-C} features \cite{Chowdhury1998}. More recent work unveiled new mechanisms to improve the photocatalytic performance of graphitic carbon nitrides to extend visible light absorption \cite{jiang2024infrared}. The coupling of infrared irradiation with lattice vibrations excites specific lattice modes, which results in the corrugation of tri-s-triazine subunit cells, activating vibration modes that may allow $\text{n}\rightarrow\pi^*$ transitions. These allowed transitions can thus enable much longer photocatalytic lifetimes while also producing \ce{H2} more efficiently \cite{jiang2024infrared}. More dedicated calculations of accurate diffusion barriers and reaction mechanisms will be needed to understand the catalytic and chemical reactivity of hydrogen onto these surfaces. Overall, future studies of the corrugation of such structures extending to other 2D-CNs will also be needed to improve industrial photocatalytic performances and, possibly, their putative photochemical evolution in conditions relevant to the ISM.\\

% Give thorough comparison with Sekine et al paper, maybe best/closer expt analogue to date
% Foley et al. gas phase ion-neutral chem
% Duley et al. grain aggregates: source of COMS?
% Rodil et al. IR of 2dcn films
% Fillion et al. review paper
% Raman IR a-cn
% Herbst et al review
% Suter et al. sensitivity to synthesis techniques and elec properties

%  \[ A = \pi r^2 \]

% \begin{equation}
%   \frac{\gamma}{\epsilon x} r^2 = 2r
% \end{equation}

% You can also put lists into the text. You can have bulleted or numbered lists of almost any kind. 
% The \texttt{mhchem} package can also be used so that formulae are easy to input: \texttt{\textbackslash ce\{H2SO4\}} gives \ce{H2SO4}. 

% For footnotes in the main text of the article please number the footnotes to avoid duplicate symbols. \textit{e.g.}\ \texttt{\textbackslash footnote[num]\{your text\}}. The corresponding author $\ast$ counts as footnote 1, Supplementary Information (SI) as footnote 2, \textit{e.g.}\ if there is no SI, please start at [num]=[2], if SI is cited in the title please start at [num]=[3] \textit{etc.} Please also cite the SI within the main body of the text using \dag.

\section*{Conclusions}

It is widely accepted that the formation of molecular species such as \ce{H2} in the ISM proceeds predominantly via heterogeneous surface chemistry on carbonaceous and silicate-like grains. Nevertheless, our understanding of the detailed compositional and physico-chemical properties of interstellar dust particles remains incomplete, as does our knowledge of the mechanisms responsible for the synthesis of specific proto-organic molecules detected in the ISM and in meteoritic materials. Diffuse interstellar clouds, where an array of organic molecules have been detected, provide an important case study to investigate gas/solid chemistry and the properties of new theoretical molecular structures. Graphitic surfaces, for example, have recently been considered as a putative system involved in molecular hydrogen formation in diffuse clouds. Circumstellar regions are also likely to harbor carbon-rich materials and (N)PAHs. 

Here, we introduced graphitic-like 2D carbon nitride nanosheets as a possible molecular class of relevance to astrochemistry, given recent experimental studies revealing that nitrogen-enriched crystalline structures may form under the low temperatures of the ISM. These nitrogen-rich structures possess unique electronic properties with a wide range of technological applications such as water photolysis, semiconductors, and new photovoltaic devices. We propose in this study that their physicochemical importance may also extend to ISM environments. As such, we conducted DFT calculations to investigate the adsorption characteristics of H onto the following nanosheets: \ce{C2N1}, \ce{C3N1}, \ce{C3N2}, \ce{C3N4}, \ce{C4N3}, \ce{C6N6}, \ce{C6N8}, \ce{C9N4}, and \ce{C9N7}. Multiple H/2D-CN chemisorption sites are identified over \ce{C-C} bonds, above C and N atoms, and at macropore center locations. We find 2D-CNs that are most favourable (\ce{C3N2}, \ce{C3N4}, \ce{C4N3}, and \ce{C9N7}), intermediately favourable (\ce{C2N1}, \ce{C6N6}, and \ce{C9N4}), and non favourable (\ce{C3N1}, \ce{C6N8}) to H adsorption. Graphitic-like carbon nitrides exhibit similar moieties to those of proto-organic molecules detected in the ISM, thereby suggesting their possible involvement in degradation pathways and/or nitrogen-rich chemical growth processes of astrochemical interest. Given their unique electronic and physicochemical characteristics, these structures may thus provide an important platform for H adsorption and in further organic growth catalysis. Future experimental and theoretical studies are needed to elucidate the possible implications of carbon nitrides on the chemical evolution of interstellar organic species, particularly in environments of potential coexistence with graphitic and nitrogen-rich molecules within interstellar clouds and planetary atmospheres.\\

\section*{Author contributions}
%%%j'ai utilisé les mots clefs du lien ci-dessous
D. D.: conceptualization, supervision, figures, writing -- original draft, writing -- review \& editing. 
P. G.: DFT calculations, figures, writing -- original draft.
R. P.: DFT calculations, writing -- original draft.\\

% All authors have given approval to the nal version of the manuscript.
% We strongly encourage authors to include author contributions and recommend using \href{https://casrai.org/credit/}{CRediT} for standardised contribution descriptions. Please refer to our general \href{https://www.rsc.org/journals-books-databases/journal-authors-reviewers/author-responsibilities/}{author guidelines} for more information about authorship.

\section*{Conflicts of interest}
There are no conflicts to declare.\\

\section*{Data availability}

The data supporting this article is available in the Supplementary Information.\\

% A data availability statement (DAS) is required to be submitted alongside all articles. Please read our \href{https://www.rsc.org/journals-books-databases/author-and-reviewer-hub/authors-information/prepare-and-format/data-sharing/#dataavailabilitystatements}{full guidance on data availability statements} for more details and examples of suitable statements you can use.

\section*{Acknowledgements}
The authors are grateful to two anonymous reviewers for their constructive suggestions.
The authors would also like to acknowledge the High Performance Computing Center of the University of Strasbourg for supporting this work by providing scientific support and access to computing resources. Part of the computing resources were funded by the Equipex Equip@Meso project (Programme Investissements d’Avenir) and the CPER Alsacalcul/Big Data. Support from NASA SMD is acknowledged. The authors gratefully acknowledge the Gauss Centre for Supercomputing e.V. (www.gauss-centre.eu) for funding this project by providing computing time on the GCS Supercomputer SuperMUC-NG at Leibniz Supercomputing Centre (www.lrz.de). The authors gratefully acknowledge the computing time made available to them on the high-performance computer Otus at the NHR Center Paderborn Center for Parallel Computing (PC2). This center is jointly supported by the Federal Ministry of Research, Technology and Space and the state governments participating in the National High-Performance Computing (NHR) joint funding program.  We wish to thank Vincent Nyamori, Akant Vats, Christiaan Boersma, Alexandros Maragkoudakis, and Partha Bera for enlightening discussions. We also gratefully acknowledge the IMAMPC 25 workshop (Caen, France) for fostering the collaboration that contributed to the origin and development of this work. The authors are grateful to A. Bafekry for kindly providing the Crystallographic Information Files (CIFs) from his published works.\\

%%%END OF MAIN TEXT%%%


\balance

\renewcommand\refname{References}

%%%REFERENCES%%%
\bibliography{full_ref_dec,Biblio_manual} %You need to replace "rsc" on this line with the name of your .bib file
\bibliographystyle{rsc} %the RSC's .bst file
\end{document}